\documentclass[aps,10pt,prl,reprint,twocolumn,showpacs,superscriptaddress,noeprint]{revtex4-1}

\usepackage{amsfonts,amsmath,amssymb}
\usepackage[margin=20mm,inner=20mm,marginpar=10mm]{geometry}
\usepackage{graphicx, epstopdf}
\usepackage{microtype}
\usepackage[figuresright]{rotating}
\usepackage{setspace}
\usepackage{textcomp}
\usepackage{times}
\usepackage[normalem]{ulem}
\usepackage[abs]{overpic}
\usepackage{color}

\usepackage[T1]{fontenc}
\usepackage[utf8]{inputenc}
\usepackage{physics}
\usepackage{bm}

\newlength{\figwidth}
\begin{document}

\title{Bragg diffraction of large organic molecules}

\author{Christian Brand}
\affiliation{University of Vienna, Faculty of Physics, Boltzmanngasse 5, A-1090 Vienna, Austria}
\affiliation{German Aerospace Center (DLR), Institute of Quantum Technologies, Söflinger Stra\ss e 100, 89077 Ulm, Germany}

\author{Filip Kia{\l}ka}
\affiliation{University of Vienna, Faculty of Physics, Boltzmanngasse 5, A-1090 Vienna, Austria}
\affiliation{Faculty of Physics, University of Duisburg-Essen, Lotharstra\ss e 1, 47048 Duisburg, Germany}

\author{Stephan Troyer}
\affiliation{University of Vienna, Faculty of Physics, Boltzmanngasse 5, A-1090 Vienna, Austria}

\author{Christian Knobloch}
\affiliation{University of Vienna, Faculty of Physics, Boltzmanngasse 5, A-1090 Vienna, Austria}

\author{Ksenija Simonovi\'c}
\affiliation{University of Vienna, Faculty of Physics, Boltzmanngasse 5, A-1090 Vienna, Austria}

\author{Benjamin A. Stickler}
\affiliation{Faculty of Physics, University of Duisburg-Essen, Lotharstra\ss e 1, 47048 Duisburg, Germany}
\affiliation{QOLS, Blackett Laboratory, Imperial College London, SW7 2AZ London, United Kingdom}

\author{Klaus Hornberger}
\affiliation{Faculty of Physics, University of Duisburg-Essen, Lotharstra\ss e 1, 47048 Duisburg, Germany}

\author{Markus Arndt}
\email{markus.arndt@univie.ac.at}
\affiliation{University of Vienna, Faculty of Physics, Boltzmanngasse 5, A-1090 Vienna, Austria}

\date{\today}

\begin{abstract}
	We demonstrate Bragg diffraction of the antibiotic ciprofloxacin and the dye molecule phthalocyanine at a thick optical grating.
	The observed patterns show a single dominant diffraction order with the expected dependence on the incidence angle as well as oscillating population transfer between the undiffracted and diffracted beams.
	We achieve an equal-amplitude splitting of $14 \hbar k$ (photon momenta) and maximum momentum transfer of $18 \hbar k$.
	This paves the way for efficient, large-momentum beam splitters and mirrors for hot and complex molecules.
\end{abstract}

\maketitle

{\it Introduction ---}
	Matter-wave diffraction and interference have numerous applications across the natural sciences.
	Electron and neutron diffraction are key techniques in condensed-matter physics and materials science~\cite{VanHoveWeinbergChan1986,Dachs_NeutronDiffraction}, while atom interferometers are utilized in tests of fundamental physics, as well as for measuring physical constants and inertial forces~\cite{Cronin2009,Tino_AtomInterferometry}.
	Extending matter-wave interference experiments to large molecules enabled quantum-assisted studies of molecular properties~\cite{Tuexen2010,Eibenberger_PRL112_250402} as well as the interference of biomolecules~\cite{ArndtACIE2017,ArndtNC2020} and particles with masses beyond $25000 \: \text{u}$~\cite{ArndtNP2019}.

	One of the major techniques used in matter-wave interferometry is Bragg diffraction.
	It employs thick gratings~\cite{Gaylord_ApplOpt20_3271} to coherently scatter the impinging particles into a single diffraction order.
	This allows for the realization of efficient matter-wave mirrors and beam splitters~\cite{Mueller_PRL100_180405,Chiow_PRL107_130403}.
	Bragg diffraction stands in contrast to Raman-Nath diffraction at thin gratings~\cite{Gould1986,Nairz2001,BatelaanN2001}, which produces several diffraction orders arranged symmetrically around the incoming particle beam.
	Bragg diffraction was first demonstrated for neutrons~\cite{Mitchell_PhysRev50_486} and later for atoms~\cite{Martin_PRL60_515}, Bose-Einstein condensates~\cite{PhillipsPRL1999}, electrons~\cite{BatelaanPRL2002}, and diatomic molecules~\cite{Abo-Shaeer_PRL94_040405}. 

\begin{figure}[t!]
	\includegraphics[width=\linewidth]{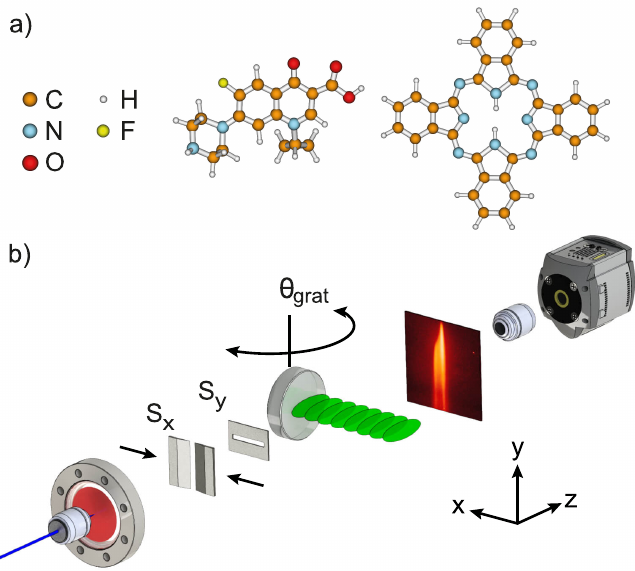}
	\caption{
		a) The experiments are performed with the antibiotic ciprofloxacin (left) and the organic dye phthalocyanine (right).
		b) A thermal beam of molecules is produced by micro-evaporation and collimated vertically (S$_x$) and horizontally (S$_y$).
		After $1.5 \: \mathrm{m}$ of free flight the molecules are diffracted at a thick laser grating created by retro-reflecting a 532~nm laser at a highly reflective mirror.
		The angle of the mirror with respect to the molecular beam $\theta_{\rm grat}$ can be controlled with µrad precision.
		The molecular diffraction pattern is recorded after further $0.57 \: \mathrm{m}$ of free flight by laser-induced fluorescence microscopy.
	}
	\label{fig:Setup}
\end{figure}

	Here, we report on the first Bragg diffraction of complex organic molecules.
	We show that the antibiotic ciprofloxacin and the dye molecule phthalocyanine [see Fig.~\ref{fig:Setup}a)] can be reliably diffracted, despite being in a highly excited rotational state and possessing more than one hundred vibrational degrees of freedom thermalized at 700--1000 K.
	This is an important step towards efficient coherent manipulation of functional, hot, and polar molecules.

{\it Experimental Setup ---}
	The experimental setup is shown in Fig.~\ref{fig:Setup}b): the molecules are evaporated by a focused laser beam, diffracted at a thick optical grating and collected on a quartz slide at the end of the vacuum chamber.
	In detail, a thin film of molecules is evaporated from the entrance window of a vacuum chamber by focusing a 420~nm laser beam down to a waist of $1.3(1) \: \text{µm}$.
	We have used mass spectrometry to verify that molecular fragmentation can be neglected in the evaporation process~\footnote{
		For details see Supplemental Material and Refs.~\cite{PhD_ChristianKnobloch,Juffmann2012,ArndtSA2017,NoriCPC2012,NoriCPC2013} included therein.
	}\nocite{PhD_ChristianKnobloch,Juffmann2012,ArndtSA2017,NoriCPC2012,NoriCPC2013}.
	After 1505~mm of free flight the molecular beam is transversely collimated with a piezo-controlled slit $\mathrm{S}_{x}$, which we set to about $14 \: \text{µm}$.
	After an additional 35~mm, the molecules are diffracted at a standing light wave, realized by retro-reflecting a laser beam with wavelength $\lambda = 532 \: \mathrm{nm}$, power $P \le  14.6(2) \: \mathrm{W}$, and waist along the flight direction $w_z=7.04(5) \: \text{mm}$.
	The waist along $y$ at the position of the molecular beam is set to $w_y=55$--$65 \: \text{µm}$ as measured with a fiber-based beam profiler~\cite{Brand_OptEx2019}.
	The angle between the mirror surface and the molecular beam $\theta_{\rm grat}$ is determined with an accuracy of about $5 \: \text{µrad}$.
	This is achieved by finding the zero-incidence position of the actively stabilized piezo mirror mount and tilting it by the desired $\theta_\text{grat}$ before each run.
	Free fall in the gravitational field leads to a vertical dispersion of the molecular velocity at the detector surface.
	To ensure good velocity separation, a vertical slit $\mathrm{S}_y$ with an opening of $25 \: \text{µm}$ is placed about 20~mm in front of the grating.
	The slit is aligned with respect to the grating with an accuracy of 10~µm using the fiber-based profiler.
	The molecular diffraction pattern is collected on a quartz plate, 570~mm behind the grating and imaged using fluorescence microscopy~\cite{Note1}.
	The experiment is conducted at a pressure below $10^{-7}$~mbar to avoid collisional decoherence.

	Ciprofloxacin is a polar biomolecule with a mass of $m=331 \: \text{u}$ and and a negligible absorption cross section of $\sigma_{\rm abs}\ll 10^{-18}$~cm$^2$ for $\lambda \ge 400$~nm~\cite{LIN20141}.
	It interacts with the light grating via its optical polarizability volume $\alpha'_{532} = 38.9$~\AA$^3$, which we calculated for the ground state geometry at the PBE0/Def2TZVPP level.
	Phthalocyanine is a non-polar dye molecule with a mass of $m=515$~u and a static polarizability volume $\alpha' =101$~\AA$^3$~\cite{Ramprasad_ApplPhysLett88_222903}.
	In contrast to ciprofloxacin, it has a non-negligible absorption cross section of $\sigma_{\rm abs} = 9 \times 10^{-18}$~cm$^2$ at 532 nm~\cite{Du_PhotochemPhotobiol68_141}.
	This allows us to observe the effect of absorption of the grating photons on the diffraction process.

{\it Theoretical model ---}
	Atomic and molecular Bragg diffraction follows from the induced dipole interaction of a polarizable point particle with a thick light grating.
	The particle moves initially with a velocity $\bm{v} = \left(v_x, v_y, v_z\right)$, where $v_z \gg v_x, v_y$.
	Since the forward momentum $m v_z$ and the kinetic energy $m \bm{v}^2/2$ are much bigger than, respectively,  the photon momentum $\hbar k = 2 \pi \hbar / \lambda$ and the potential depth, the motion in the $z$-direction is virtually unchanged by the grating and can be treated classically.
	The same can be assumed about the $y$-motion.
	Furthermore, the high $v_z$ allows us to neglect the free fall during the particle's passage through the grating.
	All this reduces the problem to the 1D dynamics along the $x$-axis.

	In a frame moving with the velocity $v_x$ the particle is initially at rest while the grating is moving.
	The Hamiltonian can then be written as
	\begin{equation} \label{eq:Hamiltonian}
		\hat{H} = - \frac{\hbar^2}{2m} \frac{\partial^2}{\partial x^2} - V(t) \cos^2 \! \big[k(x + v_x t)\big],
	\end{equation}
	where $V(t) = 16 P \alpha' / (c w_z w_y) \exp(- 2 v_z^2 t^2 / w_z^2)$.
	The time-dependent Schrödinger equation $i \hbar \partial_t \psi(t,x) = \hat{H} \psi(t,x)$ can be solved using the ansatz
	\begin{align}
		\phi\qty(t,x) &\equiv \exp(-\frac{i}{2 \hbar} \int_{-\infty}^{t} \mathrm{d}t' \, V(t')) \psi(t,x + \pi / 2k) \\
		&= \sum_{j=-\infty}^{\infty} c_j(t) \mathrm{e}^{i k j x / n}, \label{eq:ansatz}
	\end{align}
	where $n \in \mathbb{N}$ is an arbitrary integer which determines the spacing between the basis states.
	For plane-wave illumination an $n = 1$ ansatz is sufficient; for numerical simulation with finite collimation, however, $n \gg 1$ is necessary.
	Substituting Eq.~\eqref{eq:ansatz} into the Schrödinger equation yields the Raman-Nath equations~\cite{NathPIASA1936}
	\begin{equation} \label{eq:Raman-Nath}
		i c^\prime_j = \left(\frac{j}{n}\right)^2 c_j + \frac{\gamma}{4} \mathrm{e}^{-2 \tau^2 / \sigma^2} \qty[c_{j-2n} \mathrm{e}^{4 i p_\text{tr} \tau} + c_{j+2n} \mathrm{e}^{-4 i p_\text{tr} \tau}],
	\end{equation}
	where a prime denotes a derivative over $\tau = \omega_r t$, $\omega_r = \hbar k^2 / 2m$, and
	\begin{equation}
		\gamma = \frac{V(0)}{\hbar \omega_r}, \quad
		\sigma = \frac{w_z \omega_r}{v_z}, \quad
		p_\text{tr} = \frac{m v_x}{\hbar k}.
	\end{equation}
	These correspond to dimensionless grating strength, interaction time, and momentum of the incident particle.
	The interaction time parameter is close to the ratio of the grating waist radius $w_z$ and the characteristic length scale of near-field diffraction, the Talbot length $L_\text{T} = \lambda^2 m v / 4 h$~\cite{Talbot1836}, for we have $\sigma = \pi w_z / 4 L_\text{T}$.

	The Raman-Nath equations have approximate, closed-form solutions in the short-interaction and in the weak-potential limits.
	The thin-grating (or Raman-Nath) approximation amounts to dropping the kinetic term in Eq.~\eqref{eq:Hamiltonian}, which is possible when the motion of the particle inside the grating can be neglected.
	This requires $\sigma p_\text{tr} \ll 1$ and $\sigma \sqrt{\gamma} \ll 1$.
	In this regime the diffraction pattern is symmetric and independent of the incidence angle.
	The weak-grating (or Bragg) approximation amounts to the adiabatic ellimination of all but two of the Raman-Nath equations.
	This is possible when the depth of the grating potential is small compared to the recoil energy, such that only transfer to the Bragg-reflected state is allowed by energy conservation.
	For $p_\text{tr} > 1$ this is the case when $\gamma \ll 8(p_\text{tr}-1)$~\cite{ChuPRA2008}.
	In this regime, the interaction time necessary to achieve high-order diffraction grows like a factorial $\Gamma (p_\text{tr})$, as the particle has to tunnel through increasingly many energy-forbidden states.

	When the above approximations cannot be used, the solution can be obtained either via adiabatic expansion~\cite{ChuPRA2008} or numerically.
	In our experiments with ciprofloxacin $\gamma \simeq 55$, $\sigma \simeq 0.38$, and $p_\text{tr} \simeq 5$ (at $250 \: \text{m/s}$), which lies in this intermediate regime.
	We resort to numerical solution, since the convergence of the adiabatic expansion is slow.
	We note that in the intermediate regime both Raman-Nath-like and Bragg-like (also called quasi-Bragg~\cite{ChuPRA2008}) diffraction can occur, depending on the intensity profile and the thickness of the grating~\cite{RempeAPB1999}.
	We use a smooth Gaussian profile with sufficient thickness to demonstrate Bragg-like diffraction.
	The latter differs from diffraction in the weak-potential limit in that the intermediate diffraction orders are populated during the transit through the grating.
	This can lead to losses if the interaction time and strength are not optimally chosen.
	Finally, we note that classical dynamics of particles in sinusoidal potentials can give rise to analogous beam-splitting behavior~\cite{KaivolaPRL2000}; however, a quantum model is generalizable and appropriate in the absence of plausible decoherence channels.

\begin{figure}[t!]
	\includegraphics[width=\columnwidth]{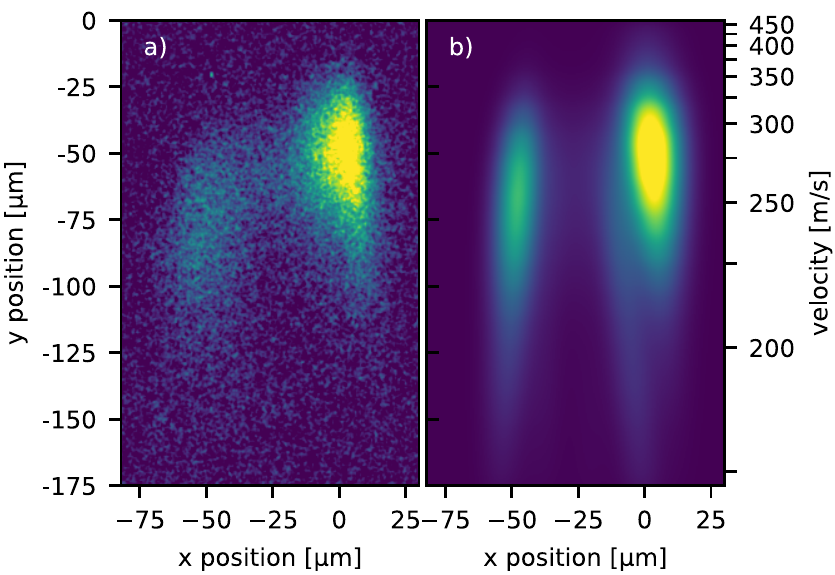}
	\caption{
		False color image of the experimental (a) and simulated (b) Bragg diffraction pattern of the antibiotic ciprofloxacin.
		The laser grating waists are $w_z =7.04(5) \: \text{mm}$, $w_y= 55(5) \: \text{µm}$, and the collimation slit is set to $14 \: \text{µm}$.
	}
	\label{fig:Bragg_cipro}
\end{figure}

{\it Diffraction of ciprofloxacin ---}
	In Fig.~\ref{fig:Bragg_cipro}a) we show the pattern obtained by diffracting ciprofloxacin molecules at an incidence angle $\theta_{\rm grat} = - 43(5) \: \text{µrad}$.
	The $y$ position in the image determines the forward velocity of the particles, which in turn determines the transverse momentum $p_\text{tr}$ and the particle-grating interaction time $\sigma$.

	For velocities above $300 \: \text{m/s}$ the interaction time is short compared to the inverse of the characteristic frequencies of the resonant Bragg transitions, and thus no diffraction occurs.
	As the characteristic frequencies increase sharply with decreasing $p_\text{tr}$~\cite{ChuPRA2008}, a relatively sudden onset of diffraction is observed at about $300 \: \text{m/s}$.
	In the $300$--$150 \: \text{m/s}$ velocity range the molecules become consecutively resonant with the 6\textsuperscript{th}--4\textsuperscript{th} Bragg transition.
	The expected momentum transfer in a Bragg transition of order $l$ is $2 l \hbar k  \propto v_z$.
	Since the flight time between the grating and the detector is inversely proportional to the forward velocity, we expect an approximately constant separation between the diffracted and the undiffracted beams.
	The slight bend in the diffracted beam results from the fact that the 6\textsuperscript{th} order transition is dominant and thus contributes also at non-resonant velocities.

	As the Bragg condition is relaxed by the limited interaction time, we observe no diffraction-free regions in between the resonances.
	Nevertheless, the appearance of a single diffracted beam and the asymmetry of the pattern help distinguish the observed phenomenon from stochastic photon absorption or Raman-Nath diffraction.
	We finally note that at $v_z \simeq 210$~m/s the amplitude of the diffracted beam matches that of the undiffracted one, demonstrating a $10 \hbar k$ equal-amplitude beam splitter.

	Numerical simulation using the Raman-Nath equations~\eqref{eq:Raman-Nath} (Fig.~\ref{fig:Bragg_cipro}b) qualitatively reproduces the observed pattern~\cite{Note1}.
	The experimental and the simulated images are vertically aligned by matching the heights at which the diffracted peaks reach half of their maximal intensities.
	This determines the most probable velocity in the molecular beam of about 250 m/s.

	\begin{figure}[t!]
		\includegraphics[width=\columnwidth]{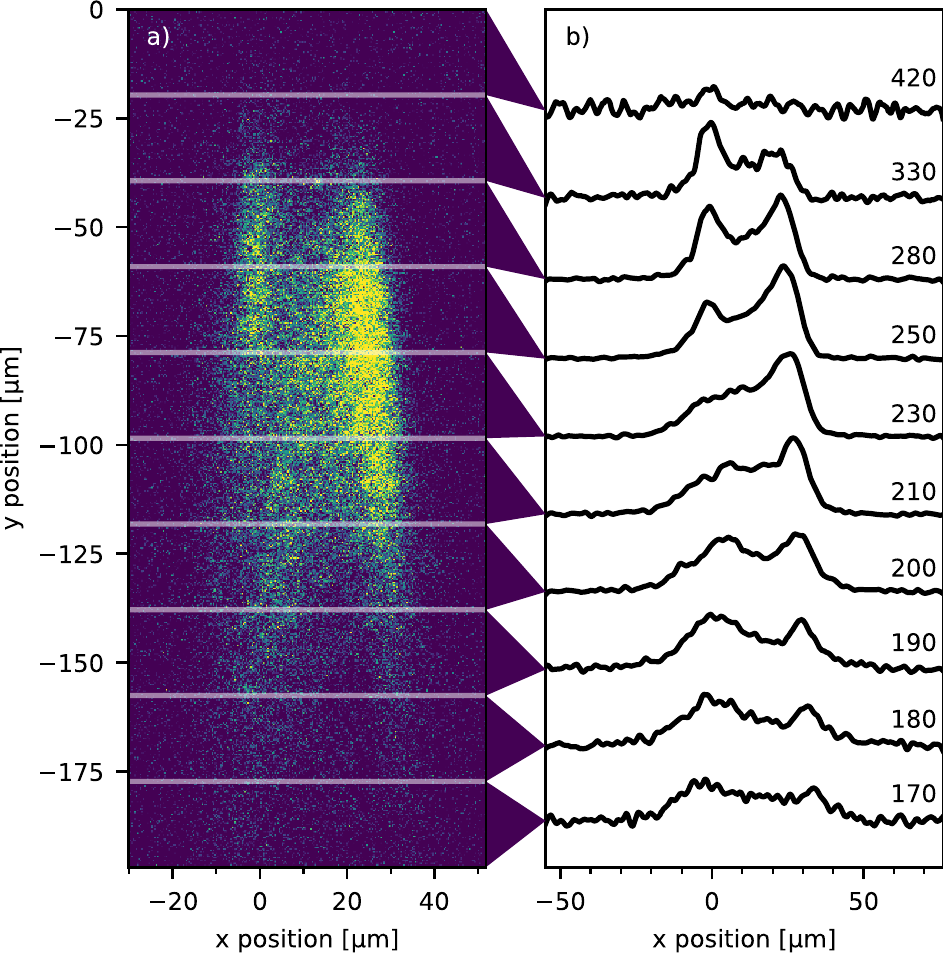}
		\caption{
			Bragg diffraction pattern of the organic dye molecule phthalocyanine at an incidence angle $\theta_\text{grat} = 5(5) \: \text{µrad}$.
			Panel (a) shows the false color diffraction image.
			Panel (b) shows the averages of $20 \: \text{µm}$ high stripes, smoothed with median and Savitzky–Golay~\cite{Savitzky_AC36_1627} filters, and annotated with their corresponding velocities in $\text{m/s}$.
			The velocities are determined by comparison with a diffraction pattern produced by a material grating~\cite{Note1}.
			The laser grating waist for this measurement is $w_y=57(3) \: \text{µm}$ and the collimation slit width is $11.5 \: \text{µm}$.
		}
		\label{fig:pch2_pendelloesung}
	\end{figure}

{\it Diffraction of phthalocyanine ---}
	To explore the universality of molecular Bragg diffraction and its robustness to absorption, we switch to the dye molecule phthalocyanine.
	We quantify the absorption by setting $\theta_\text{grat}$ to an angle for which we do not expect diffraction and observing the broadening of the molecular beam.
	From the width of the beam we infer that on average one photon is absorbed inside the grating~\cite{Note1}.
	Despite the absorption, we obtain diffraction images of phthalocyanine, which are qualitatively similar to those of ciprofloxacin [see Fig.~\ref{fig:pch2_pendelloesung}a)].
	The images exhibit oscillating population transfer [see Fig.~\ref{fig:pch2_pendelloesung}b)] reminiscent of the Pendellösung oscillations predicted by the theory of weak-potential Bragg diffraction and demonstrated with neutrons~\cite{Shull_PhysRevLett21_1585} and atoms~\cite{Oberthaler_PRA60_456}.
	Similar oscillations can be seen in the power dependence of the diffraction patterns~\cite{Note1}.

\begin{figure}[t!]
	\includegraphics[width=\columnwidth]{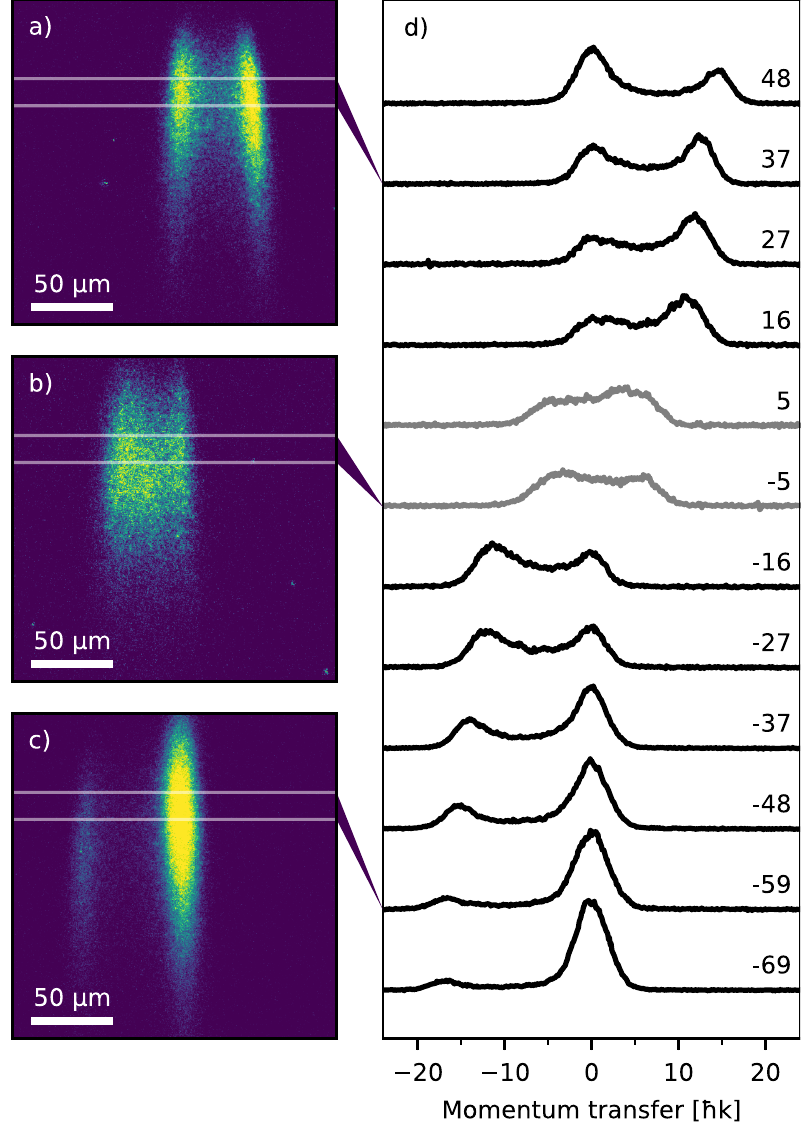}
	\caption{
		Angular dependence of Bragg diffraction of the dye molecule phthalocyanine.
		Panels (a)--(c) show diffraction images for the incidence angles $48$ (a), $-5$ (b) and $-69 \: \text{µrad}$ (c).
		The images are $197$ by $197 \: \text{µm}$ and the scale bars are $50 \: \text{µm}$ long.
		Panel (d) shows the integrated intensity profiles for the incidence angle varying from $-69$ to $48 \: \text{µrad}$ in steps of $10 \: \text{µrad}$.
		The profiles are averages of $16 \: \text{µm}$ high stripes of the diffraction images corresponding to a velocity range of $234$ to $255 \: \mathrm{m/s}$.
		The curves are horizontally aligned to center the undiffracted beam (which is the right peak for negative incidence and the left peak for positive incidence).
		For $\pm 5 \: \text{µrad}$ we observe diffraction to both sides of the initial beam and hence align the traces with respect to their center of gravity.
		The laser grating waist for this measurement is $w_y=65(5) \: \text{µm}$ and the collimation slit width is $14.8 \: \text{µm}$.
	}
	\label{fig:RockingCurve}
\end{figure}

	To investigate the dependence of Bragg diffraction on the incidence angle, we record a series of diffraction images in which we vary $\theta_\text{grat}$ (see Fig.~\ref{fig:RockingCurve}).
	In agreement with the expectations, we find the molecules diffracted to either side of the incoming beam, depending on the sign of the incidence angle.
	Similarly as for ciprofloxacin, the diffracted molecules form a slanted stripe indicating a single dominant transition.
	This transition is broadened by the $12 \: \text{µrad}$ collimation of the molecular beam~\cite{Note1}, which results in deviations from specular reflection seen in Figs.~\ref{fig:pch2_pendelloesung} and~\ref{fig:RockingCurve}.
	The highest momentum transfer recorded was $18 \hbar k$ with an efficiency of 10\% [Fig.~\ref{fig:RockingCurve}c)], and equal-amplitude splitting was realized for a momentum separation of $14 \hbar k$ [Fig.~\ref{fig:RockingCurve}a)].

{\it Discussion and outlook ---}
	We have demonstrated Bragg diffraction for the complex organic molecules phthalocyanine and ciprofloxacin.
	As our data is in qualitative agreement with a simple polarizable-point-particle model, we expect that this technique can be applied without modification to any molecule of comparable size and absorption cross section.
	That is irrespective of the details of its electronic structure, dipole moment, etc.
	We have demonstrated a balanced beam splitter with a momentum separation of $14 \hbar k$, which is to the best of our knowledge the largest equal-amplitude splitting demonstrated for molecules using optical gratings.
	Although with sufficient laser power similar or even greater splitting could be achieved with a thin optical grating, this would typically reduce the particle flux by a factor of 10 as only two of the many populated output beams have to be selected.
	The same problem applies to mechanical gratings, which additionally are incompatible with polar molecules due to rotational averaging.

	Further development should increase the particle-grating interaction time in order to decrease losses and sharpen the Bragg resonances.
	A promising approach to achieve this is slowing the molecules using buffer gas cells~\cite{DoyleCPC2014}.
	This could ultimately allow for Mach-Zehnder interferometry with large molecules.
	The possibility to selectively address the arms in such a setup would, in turn, enable new interference schemes utilizing the molecules' chirality, conformation, and possibly entanglement between the molecules' internal and external degrees of freedom.
	Efficient Bragg diffraction could also enable pulsed Bloch oscillation beam splitters to realize even larger momentum transfers~\cite{SalomonPRL1996}.

{\it Acknowledgements ---}
	This project has received funding from the European Research Council (ERC) under the European Union's Horizon 2020 research and innovation program (Grant Nr. 320694) and the Austrian Science Fund (FWF) within project P-30176 and W1210-N25.
	We thank the Fetzer Pioneers' Fund (project P\#2018-1) for financial support, Armin Shayeghi for calculating the polarizability of ciprofloxacin and Martin Fally for fruitful discussions.
	B.A.S. acknowledges funding from the European Union’s Horizon 2020 research and innovation programme under the Marie Skłodowska-Curie grant agreement No. 841040.


\begin{thebibliography}{40}%
\makeatletter
\providecommand \@ifxundefined [1]{%
 \@ifx{#1\undefined}
}%
\providecommand \@ifnum [1]{%
 \ifnum #1\expandafter \@firstoftwo
 \else \expandafter \@secondoftwo
 \fi
}%
\providecommand \@ifx [1]{%
 \ifx #1\expandafter \@firstoftwo
 \else \expandafter \@secondoftwo
 \fi
}%
\providecommand \natexlab [1]{#1}%
\providecommand \enquote  [1]{``#1''}%
\providecommand \bibnamefont  [1]{#1}%
\providecommand \bibfnamefont [1]{#1}%
\providecommand \citenamefont [1]{#1}%
\providecommand \href@noop [0]{\@secondoftwo}%
\providecommand \href [0]{\begingroup \@sanitize@url \@href}%
\providecommand \@href[1]{\@@startlink{#1}\@@href}%
\providecommand \@@href[1]{\endgroup#1\@@endlink}%
\providecommand \@sanitize@url [0]{\catcode `\\12\catcode `\$12\catcode
  `\&12\catcode `\#12\catcode `\^12\catcode `\_12\catcode `\%12\relax}%
\providecommand \@@startlink[1]{}%
\providecommand \@@endlink[0]{}%
\providecommand \url  [0]{\begingroup\@sanitize@url \@url }%
\providecommand \@url [1]{\endgroup\@href {#1}{\urlprefix }}%
\providecommand \urlprefix  [0]{URL }%
\providecommand \Eprint [0]{\href }%
\providecommand \doibase [0]{https://doi.org/}%
\providecommand \selectlanguage [0]{\@gobble}%
\providecommand \bibinfo  [0]{\@secondoftwo}%
\providecommand \bibfield  [0]{\@secondoftwo}%
\providecommand \translation [1]{[#1]}%
\providecommand \BibitemOpen [0]{}%
\providecommand \bibitemStop [0]{}%
\providecommand \bibitemNoStop [0]{.\EOS\space}%
\providecommand \EOS [0]{\spacefactor3000\relax}%
\providecommand \BibitemShut  [1]{\csname bibitem#1\endcsname}%
\let\auto@bib@innerbib\@empty
\bibitem [{\citenamefont {VanHove}\ \emph {et~al.}(1986)\citenamefont
  {VanHove}, \citenamefont {Weinberg},\ and\ \citenamefont
  {Chan}}]{VanHoveWeinbergChan1986}%
  \BibitemOpen
  \bibfield  {author} {\bibinfo {author} {\bibfnamefont {M.}~\bibnamefont
  {VanHove}}, \bibinfo {author} {\bibfnamefont {W.}~\bibnamefont {Weinberg}},\
  and\ \bibinfo {author} {\bibfnamefont {C.}~\bibnamefont {Chan}},\ }\href@noop
  {} {\emph {\bibinfo {title} {Low-Energy Electron Diffraction: Experiment,
  Theory and Surface Structure Determination}}},\ \bibinfo {series} {Springer
  Series in Surface Sciences}, Vol.~\bibinfo {volume} {6}\ (\bibinfo
  {publisher} {Springer-Verlag Berlin Heidelberg},\ \bibinfo {year}
  {1986})\BibitemShut {NoStop}%
\bibitem [{\citenamefont {Dachs}(1978)}]{Dachs_NeutronDiffraction}%
  \BibitemOpen
  \bibinfo {editor} {\bibfnamefont {H.}~\bibnamefont {Dachs}},\ ed.,\
  \href@noop {} {\emph {\bibinfo {title} {Neutron Diffraction}}},\ Topics in
  Current Physics\ (\bibinfo  {publisher} {Springer},\ \bibinfo {address}
  {Heidelberg},\ \bibinfo {year} {1978})\BibitemShut {NoStop}%
\bibitem [{\citenamefont {Cronin}\ \emph {et~al.}(2009)\citenamefont {Cronin},
  \citenamefont {Schmiedmayer},\ and\ \citenamefont {Pritchard}}]{Cronin2009}%
  \BibitemOpen
  \bibfield  {author} {\bibinfo {author} {\bibfnamefont {A.~D.}\ \bibnamefont
  {Cronin}}, \bibinfo {author} {\bibfnamefont {J.}~\bibnamefont
  {Schmiedmayer}},\ and\ \bibinfo {author} {\bibfnamefont {D.~E.}\ \bibnamefont
  {Pritchard}},\ }\bibfield  {title} {\bibinfo {title} {Optics and
  interferometry with atoms and molecules},\ }\href
  {https://doi.org/10.1103/RevModPhys.81.1051} {\bibfield  {journal} {\bibinfo
  {journal} {Rev. Mod. Phys.}\ }\textbf {\bibinfo {volume} {81}},\ \bibinfo
  {pages} {1051} (\bibinfo {year} {2009})}\BibitemShut {NoStop}%
\bibitem [{\citenamefont {Tino}\ and\ \citenamefont
  {Kasevich}(2014)}]{Tino_AtomInterferometry}%
  \BibitemOpen
  \bibinfo {editor} {\bibfnamefont {G.~M.}\ \bibnamefont {Tino}}\ and\ \bibinfo
  {editor} {\bibfnamefont {M.~A.}\ \bibnamefont {Kasevich}},\ eds.,\ \href@noop
  {} {\emph {\bibinfo {title} {Atom Interferometry}}},\ \bibinfo {series}
  {Proceedings of the International School of Physics "Enrico Fermi"}, Vol.\
  \bibinfo {volume} {188}\ (\bibinfo  {publisher} {IOS Press},\ \bibinfo {year}
  {2014})\BibitemShut {NoStop}%
\bibitem [{\citenamefont {Tüxen}\ \emph {et~al.}(2010)\citenamefont {Tüxen},
  \citenamefont {Gerlich}, \citenamefont {Eibenberger}, \citenamefont {Arndt},\
  and\ \citenamefont {Mayor}}]{Tuexen2010}%
  \BibitemOpen
  \bibfield  {author} {\bibinfo {author} {\bibfnamefont {J.}~\bibnamefont
  {Tüxen}}, \bibinfo {author} {\bibfnamefont {S.}~\bibnamefont {Gerlich}},
  \bibinfo {author} {\bibfnamefont {S.}~\bibnamefont {Eibenberger}}, \bibinfo
  {author} {\bibfnamefont {M.}~\bibnamefont {Arndt}},\ and\ \bibinfo {author}
  {\bibfnamefont {M.}~\bibnamefont {Mayor}},\ }\bibfield  {title} {\bibinfo
  {title} {Quantum interference distinguishes between constitutional isomers},\
  }\href@noop {} {\bibfield  {journal} {\bibinfo  {journal} {Chem. Comm.}\
  }\textbf {\bibinfo {volume} {46}},\ \bibinfo {pages} {4145} (\bibinfo {year}
  {2010})}\BibitemShut {NoStop}%
\bibitem [{\citenamefont {Eibenberger}\ \emph {et~al.}(2014)\citenamefont
  {Eibenberger}, \citenamefont {Cheng}, \citenamefont {Cotter},\ and\
  \citenamefont {Arndt}}]{Eibenberger_PRL112_250402}%
  \BibitemOpen
  \bibfield  {author} {\bibinfo {author} {\bibfnamefont {S.}~\bibnamefont
  {Eibenberger}}, \bibinfo {author} {\bibfnamefont {X.}~\bibnamefont {Cheng}},
  \bibinfo {author} {\bibfnamefont {J.~P.}\ \bibnamefont {Cotter}},\ and\
  \bibinfo {author} {\bibfnamefont {M.}~\bibnamefont {Arndt}},\ }\bibfield
  {title} {\bibinfo {title} {Absolute absorption cross sections from photon
  recoil in a matter-wave interferometer},\ }\href@noop {} {\bibfield
  {journal} {\bibinfo  {journal} {Phys. Rev. Lett.}\ }\textbf {\bibinfo
  {volume} {112}},\ \bibinfo {pages} {250402} (\bibinfo {year}
  {2014})}\BibitemShut {NoStop}%
\bibitem [{\citenamefont {Mairhofer}\ \emph {et~al.}(2017)\citenamefont
  {Mairhofer}, \citenamefont {Eibenberger}, \citenamefont {Cotter},
  \citenamefont {Romirer}, \citenamefont {Shayeghi},\ and\ \citenamefont
  {Arndt}}]{ArndtACIE2017}%
  \BibitemOpen
  \bibfield  {author} {\bibinfo {author} {\bibfnamefont {L.}~\bibnamefont
  {Mairhofer}}, \bibinfo {author} {\bibfnamefont {S.}~\bibnamefont
  {Eibenberger}}, \bibinfo {author} {\bibfnamefont {J.~P.}\ \bibnamefont
  {Cotter}}, \bibinfo {author} {\bibfnamefont {M.}~\bibnamefont {Romirer}},
  \bibinfo {author} {\bibfnamefont {A.}~\bibnamefont {Shayeghi}},\ and\
  \bibinfo {author} {\bibfnamefont {M.}~\bibnamefont {Arndt}},\ }\bibfield
  {title} {\bibinfo {title} {Quantum-assisted metrology of neutral vitamins in
  the gas phase},\ }\href {https://doi.org/10.1002/anie.201704916} {\bibfield
  {journal} {\bibinfo  {journal} {Angew. Chem. Int. Ed.}\ }\textbf {\bibinfo
  {volume} {56}},\ \bibinfo {pages} {10947} (\bibinfo {year}
  {2017})}\BibitemShut {NoStop}%
\bibitem [{\citenamefont {Shayeghi}\ \emph {et~al.}(2020)\citenamefont
  {Shayeghi}, \citenamefont {Rieser}, \citenamefont {Richter}, \citenamefont
  {Sezer}, \citenamefont {Rodewald}, \citenamefont {Geyer}, \citenamefont
  {Martinez},\ and\ \citenamefont {Arndt}}]{ArndtNC2020}%
  \BibitemOpen
  \bibfield  {author} {\bibinfo {author} {\bibfnamefont {A.}~\bibnamefont
  {Shayeghi}}, \bibinfo {author} {\bibfnamefont {P.}~\bibnamefont {Rieser}},
  \bibinfo {author} {\bibfnamefont {G.}~\bibnamefont {Richter}}, \bibinfo
  {author} {\bibfnamefont {U.}~\bibnamefont {Sezer}}, \bibinfo {author}
  {\bibfnamefont {J.~H.}\ \bibnamefont {Rodewald}}, \bibinfo {author}
  {\bibfnamefont {P.}~\bibnamefont {Geyer}}, \bibinfo {author} {\bibfnamefont
  {T.~J.}\ \bibnamefont {Martinez}},\ and\ \bibinfo {author} {\bibfnamefont
  {M.}~\bibnamefont {Arndt}},\ }\bibfield  {title} {\bibinfo {title}
  {Matter-wave interference of a native polypeptide},\ }\href
  {https://doi.org/10.1038/s41467-020-15280-2} {\bibfield  {journal} {\bibinfo
  {journal} {Nat. Commun.}\ }\textbf {\bibinfo {volume} {11}},\ \bibinfo
  {pages} {1447} (\bibinfo {year} {2020})}\BibitemShut {NoStop}%
\bibitem [{\citenamefont {Fein}\ \emph {et~al.}(2019)\citenamefont {Fein},
  \citenamefont {Geyer}, \citenamefont {Zwick}, \citenamefont {Kia{\l}ka},
  \citenamefont {Pedalino}, \citenamefont {Mayor}, \citenamefont {Gerlich},\
  and\ \citenamefont {Arndt}}]{ArndtNP2019}%
  \BibitemOpen
  \bibfield  {author} {\bibinfo {author} {\bibfnamefont {Y.~Y.}\ \bibnamefont
  {Fein}}, \bibinfo {author} {\bibfnamefont {P.}~\bibnamefont {Geyer}},
  \bibinfo {author} {\bibfnamefont {P.}~\bibnamefont {Zwick}}, \bibinfo
  {author} {\bibfnamefont {F.}~\bibnamefont {Kia{\l}ka}}, \bibinfo {author}
  {\bibfnamefont {S.}~\bibnamefont {Pedalino}}, \bibinfo {author}
  {\bibfnamefont {M.}~\bibnamefont {Mayor}}, \bibinfo {author} {\bibfnamefont
  {S.}~\bibnamefont {Gerlich}},\ and\ \bibinfo {author} {\bibfnamefont
  {M.}~\bibnamefont {Arndt}},\ }\bibfield  {title} {\bibinfo {title} {Quantum
  superposition of molecules beyond 25 {{kDa}}},\ }\href
  {https://doi.org/10.1038/s41567-019-0663-9} {\bibfield  {journal} {\bibinfo
  {journal} {Nat. Phys.}\ }\textbf {\bibinfo {volume} {15}},\ \bibinfo {pages}
  {1242} (\bibinfo {year} {2019})}\BibitemShut {NoStop}%
\bibitem [{\citenamefont {Gaylord}\ and\ \citenamefont
  {Moharam}(1981)}]{Gaylord_ApplOpt20_3271}%
  \BibitemOpen
  \bibfield  {author} {\bibinfo {author} {\bibfnamefont {T.~K.}\ \bibnamefont
  {Gaylord}}\ and\ \bibinfo {author} {\bibfnamefont {M.~G.}\ \bibnamefont
  {Moharam}},\ }\bibfield  {title} {\bibinfo {title} {Thin and thick gratings:
  terminology clarification},\ }\href@noop {} {\bibfield  {journal} {\bibinfo
  {journal} {Appl. Opt.}\ }\textbf {\bibinfo {volume} {20}},\ \bibinfo {pages}
  {3271} (\bibinfo {year} {1981})}\BibitemShut {NoStop}%
\bibitem [{\citenamefont {M\"uller}\ \emph {et~al.}(2008)\citenamefont
  {M\"uller}, \citenamefont {Chiow}, \citenamefont {Long}, \citenamefont
  {Herrmann},\ and\ \citenamefont {Chu}}]{Mueller_PRL100_180405}%
  \BibitemOpen
  \bibfield  {author} {\bibinfo {author} {\bibfnamefont {H.}~\bibnamefont
  {M\"uller}}, \bibinfo {author} {\bibfnamefont {S.-w.}\ \bibnamefont {Chiow}},
  \bibinfo {author} {\bibfnamefont {Q.}~\bibnamefont {Long}}, \bibinfo {author}
  {\bibfnamefont {S.}~\bibnamefont {Herrmann}},\ and\ \bibinfo {author}
  {\bibfnamefont {S.}~\bibnamefont {Chu}},\ }\bibfield  {title} {\bibinfo
  {title} {Atom interferometry with up to 24-photon-momentum-transfer beam
  splitters},\ }\href@noop {} {\bibfield  {journal} {\bibinfo  {journal} {Phys.
  Rev. Lett.}\ }\textbf {\bibinfo {volume} {100}},\ \bibinfo {pages} {180405}
  (\bibinfo {year} {2008})}\BibitemShut {NoStop}%
\bibitem [{\citenamefont {Chiow}\ \emph {et~al.}(2011)\citenamefont {Chiow},
  \citenamefont {Kovachy}, \citenamefont {Chien},\ and\ \citenamefont
  {Kasevich}}]{Chiow_PRL107_130403}%
  \BibitemOpen
  \bibfield  {author} {\bibinfo {author} {\bibfnamefont {S.-w.}\ \bibnamefont
  {Chiow}}, \bibinfo {author} {\bibfnamefont {T.}~\bibnamefont {Kovachy}},
  \bibinfo {author} {\bibfnamefont {H.-C.}\ \bibnamefont {Chien}},\ and\
  \bibinfo {author} {\bibfnamefont {M.~A.}\ \bibnamefont {Kasevich}},\
  }\bibfield  {title} {\bibinfo {title} {102 $\hbar$k large area atom
  interferometers},\ }\href@noop {} {\bibfield  {journal} {\bibinfo  {journal}
  {Phys. Rev. Lett.}\ }\textbf {\bibinfo {volume} {107}},\ \bibinfo {pages}
  {130403} (\bibinfo {year} {2011})}\BibitemShut {NoStop}%
\bibitem [{\citenamefont {Gould}\ \emph {et~al.}(1986)\citenamefont {Gould},
  \citenamefont {Ruff},\ and\ \citenamefont {Pritchard}}]{Gould1986}%
  \BibitemOpen
  \bibfield  {author} {\bibinfo {author} {\bibfnamefont {P.~L.}\ \bibnamefont
  {Gould}}, \bibinfo {author} {\bibfnamefont {G.~A.}\ \bibnamefont {Ruff}},\
  and\ \bibinfo {author} {\bibfnamefont {D.~E.}\ \bibnamefont {Pritchard}},\
  }\bibfield  {title} {\bibinfo {title} {Diffraction of atoms by light: The
  near-resonant kapitza-dirac effect},\ }\href@noop {} {\bibfield  {journal}
  {\bibinfo  {journal} {Phys. Rev. Lett.}\ }\textbf {\bibinfo {volume} {56}},\
  \bibinfo {pages} {827 } (\bibinfo {year} {1986})}\BibitemShut {NoStop}%
\bibitem [{\citenamefont {Nairz}\ \emph {et~al.}(2001)\citenamefont {Nairz},
  \citenamefont {Brezger}, \citenamefont {Arndt},\ and\ \citenamefont
  {Zeilinger}}]{Nairz2001}%
  \BibitemOpen
  \bibfield  {author} {\bibinfo {author} {\bibfnamefont {O.}~\bibnamefont
  {Nairz}}, \bibinfo {author} {\bibfnamefont {B.}~\bibnamefont {Brezger}},
  \bibinfo {author} {\bibfnamefont {M.}~\bibnamefont {Arndt}},\ and\ \bibinfo
  {author} {\bibfnamefont {A.}~\bibnamefont {Zeilinger}},\ }\bibfield  {title}
  {\bibinfo {title} {Diffraction of complex molecules by structures made of
  light},\ }\href@noop {} {\bibfield  {journal} {\bibinfo  {journal} {Phys.
  Rev. Lett.}\ }\textbf {\bibinfo {volume} {87}},\ \bibinfo {pages} {160401}
  (\bibinfo {year} {2001})}\BibitemShut {NoStop}%
\bibitem [{\citenamefont {Freimund}\ \emph {et~al.}(2001)\citenamefont
  {Freimund}, \citenamefont {Aflatooni},\ and\ \citenamefont
  {Batelaan}}]{BatelaanN2001}%
  \BibitemOpen
  \bibfield  {author} {\bibinfo {author} {\bibfnamefont {D.~L.}\ \bibnamefont
  {Freimund}}, \bibinfo {author} {\bibfnamefont {K.}~\bibnamefont
  {Aflatooni}},\ and\ \bibinfo {author} {\bibfnamefont {H.}~\bibnamefont
  {Batelaan}},\ }\bibfield  {title} {\bibinfo {title} {Observation of the
  {{Kapitza}}\textendash{{Dirac}} effect},\ }\href
  {https://doi.org/10.1038/35093065} {\bibfield  {journal} {\bibinfo  {journal}
  {Nature}\ }\textbf {\bibinfo {volume} {413}},\ \bibinfo {pages} {142}
  (\bibinfo {year} {2001})}\BibitemShut {NoStop}%
\bibitem [{\citenamefont {Mitchell}\ and\ \citenamefont
  {Powers}(1936)}]{Mitchell_PhysRev50_486}%
  \BibitemOpen
  \bibfield  {author} {\bibinfo {author} {\bibfnamefont {D.~P.}\ \bibnamefont
  {Mitchell}}\ and\ \bibinfo {author} {\bibfnamefont {P.~N.}\ \bibnamefont
  {Powers}},\ }\bibfield  {title} {\bibinfo {title} {Bragg reflection of slow
  neutrons},\ }\href@noop {} {\bibfield  {journal} {\bibinfo  {journal} {Phys.
  Rev.}\ }\textbf {\bibinfo {volume} {50}},\ \bibinfo {pages} {486} (\bibinfo
  {year} {1936})}\BibitemShut {NoStop}%
\bibitem [{\citenamefont {Martin}\ \emph {et~al.}(1988)\citenamefont {Martin},
  \citenamefont {Oldaker}, \citenamefont {Miklich},\ and\ \citenamefont
  {Pritchard}}]{Martin_PRL60_515}%
  \BibitemOpen
  \bibfield  {author} {\bibinfo {author} {\bibfnamefont {P.~J.}\ \bibnamefont
  {Martin}}, \bibinfo {author} {\bibfnamefont {B.~G.}\ \bibnamefont {Oldaker}},
  \bibinfo {author} {\bibfnamefont {A.~H.}\ \bibnamefont {Miklich}},\ and\
  \bibinfo {author} {\bibfnamefont {D.~E.}\ \bibnamefont {Pritchard}},\
  }\bibfield  {title} {\bibinfo {title} {Bragg scattering of atoms from a
  standing light wave},\ }\href@noop {} {\bibfield  {journal} {\bibinfo
  {journal} {Phys. Rev. Lett.}\ }\textbf {\bibinfo {volume} {60}},\ \bibinfo
  {pages} {515} (\bibinfo {year} {1988})}\BibitemShut {NoStop}%
\bibitem [{\citenamefont {Kozuma}\ \emph {et~al.}(1999)\citenamefont {Kozuma},
  \citenamefont {Deng}, \citenamefont {Hagley}, \citenamefont {Wen},
  \citenamefont {Lutwak}, \citenamefont {Helmerson}, \citenamefont {Rolston},\
  and\ \citenamefont {Phillips}}]{PhillipsPRL1999}%
  \BibitemOpen
  \bibfield  {author} {\bibinfo {author} {\bibfnamefont {M.}~\bibnamefont
  {Kozuma}}, \bibinfo {author} {\bibfnamefont {L.}~\bibnamefont {Deng}},
  \bibinfo {author} {\bibfnamefont {E.~W.}\ \bibnamefont {Hagley}}, \bibinfo
  {author} {\bibfnamefont {J.}~\bibnamefont {Wen}}, \bibinfo {author}
  {\bibfnamefont {R.}~\bibnamefont {Lutwak}}, \bibinfo {author} {\bibfnamefont
  {K.}~\bibnamefont {Helmerson}}, \bibinfo {author} {\bibfnamefont {S.~L.}\
  \bibnamefont {Rolston}},\ and\ \bibinfo {author} {\bibfnamefont {W.~D.}\
  \bibnamefont {Phillips}},\ }\bibfield  {title} {\bibinfo {title} {Coherent
  splitting of {Bose-Einstein} condensed atoms with optically induced {Bragg}
  diffraction},\ }\href {https://doi.org/10.1103/PhysRevLett.82.871} {\bibfield
   {journal} {\bibinfo  {journal} {Phys. Rev. Lett.}\ }\textbf {\bibinfo
  {volume} {82}},\ \bibinfo {pages} {871} (\bibinfo {year} {1999})}\BibitemShut
  {NoStop}%
\bibitem [{\citenamefont {Freimund}\ and\ \citenamefont
  {Batelaan}(2002)}]{BatelaanPRL2002}%
  \BibitemOpen
  \bibfield  {author} {\bibinfo {author} {\bibfnamefont {D.~L.}\ \bibnamefont
  {Freimund}}\ and\ \bibinfo {author} {\bibfnamefont {H.}~\bibnamefont
  {Batelaan}},\ }\bibfield  {title} {\bibinfo {title} {Bragg scattering of free
  electrons using the {Kapitza-Dirac} effect},\ }\href
  {https://doi.org/10.1103/PhysRevLett.89.283602} {\bibfield  {journal}
  {\bibinfo  {journal} {Phys. Rev. Lett.}\ }\textbf {\bibinfo {volume} {89}},\
  \bibinfo {pages} {283602} (\bibinfo {year} {2002})}\BibitemShut {NoStop}%
\bibitem [{\citenamefont {Abo-Shaeer}\ \emph {et~al.}(2005)\citenamefont
  {Abo-Shaeer}, \citenamefont {Miller}, \citenamefont {Chin}, \citenamefont
  {Xu}, \citenamefont {Mukaiyama},\ and\ \citenamefont
  {Ketterle}}]{Abo-Shaeer_PRL94_040405}%
  \BibitemOpen
  \bibfield  {author} {\bibinfo {author} {\bibfnamefont {J.~R.}\ \bibnamefont
  {Abo-Shaeer}}, \bibinfo {author} {\bibfnamefont {D.~E.}\ \bibnamefont
  {Miller}}, \bibinfo {author} {\bibfnamefont {J.~K.}\ \bibnamefont {Chin}},
  \bibinfo {author} {\bibfnamefont {K.}~\bibnamefont {Xu}}, \bibinfo {author}
  {\bibfnamefont {T.}~\bibnamefont {Mukaiyama}},\ and\ \bibinfo {author}
  {\bibfnamefont {W.}~\bibnamefont {Ketterle}},\ }\bibfield  {title} {\bibinfo
  {title} {Coherent molecular optics using ultracold sodium dimers},\
  }\href@noop {} {\bibfield  {journal} {\bibinfo  {journal} {Phys. Rev. Lett.}\
  }\textbf {\bibinfo {volume} {94}},\ \bibinfo {pages} {040405} (\bibinfo
  {year} {2005})}\BibitemShut {NoStop}%
\bibitem [{Note1()}]{Note1}%
  \BibitemOpen
  \bibinfo {note} {For details see Supplemental Material and Refs.~\cite
  {PhD_ChristianKnobloch,Juffmann2012,ArndtSA2017,NoriCPC2012,NoriCPC2013}
  included therein.}\BibitemShut {Stop}%
\bibitem [{\citenamefont {Knobloch}(2019)}]{PhD_ChristianKnobloch}%
  \BibitemOpen
  \bibfield  {author} {\bibinfo {author} {\bibfnamefont {C.}~\bibnamefont
  {Knobloch}},\ }\emph {\bibinfo {title} {Coherent matter-wave manipulation
  techniques}},\ \href@noop {} {Ph.D. thesis} (\bibinfo {year}
  {2019})\BibitemShut {NoStop}%
\bibitem [{\citenamefont {Juffmann}\ \emph {et~al.}(2012)\citenamefont
  {Juffmann}, \citenamefont {Milic}, \citenamefont {M\"ullneritsch},
  \citenamefont {Asenbaum}, \citenamefont {Tsukernik}, \citenamefont {T\"uxen},
  \citenamefont {Mayor}, \citenamefont {Cheshnovsky},\ and\ \citenamefont
  {Arndt}}]{Juffmann2012}%
  \BibitemOpen
  \bibfield  {author} {\bibinfo {author} {\bibfnamefont {T.}~\bibnamefont
  {Juffmann}}, \bibinfo {author} {\bibfnamefont {A.}~\bibnamefont {Milic}},
  \bibinfo {author} {\bibfnamefont {M.}~\bibnamefont {M\"ullneritsch}},
  \bibinfo {author} {\bibfnamefont {P.}~\bibnamefont {Asenbaum}}, \bibinfo
  {author} {\bibfnamefont {A.}~\bibnamefont {Tsukernik}}, \bibinfo {author}
  {\bibfnamefont {J.}~\bibnamefont {T\"uxen}}, \bibinfo {author} {\bibfnamefont
  {M.}~\bibnamefont {Mayor}}, \bibinfo {author} {\bibfnamefont
  {O.}~\bibnamefont {Cheshnovsky}},\ and\ \bibinfo {author} {\bibfnamefont
  {M.}~\bibnamefont {Arndt}},\ }\bibfield  {title} {\bibinfo {title} {Real-time
  single-molecule imaging of quantum interference},\ }\href@noop {} {\bibfield
  {journal} {\bibinfo  {journal} {Nat. Nanotechnol.}\ }\textbf {\bibinfo
  {volume} {7}},\ \bibinfo {pages} {297} (\bibinfo {year} {2012})}\BibitemShut
  {NoStop}%
\bibitem [{\citenamefont {Cotter}\ \emph {et~al.}(2017)\citenamefont {Cotter},
  \citenamefont {Brand}, \citenamefont {Knobloch}, \citenamefont {Lilach},
  \citenamefont {Cheshnovsky},\ and\ \citenamefont {Arndt}}]{ArndtSA2017}%
  \BibitemOpen
  \bibfield  {author} {\bibinfo {author} {\bibfnamefont {J.~P.}\ \bibnamefont
  {Cotter}}, \bibinfo {author} {\bibfnamefont {C.}~\bibnamefont {Brand}},
  \bibinfo {author} {\bibfnamefont {C.}~\bibnamefont {Knobloch}}, \bibinfo
  {author} {\bibfnamefont {Y.}~\bibnamefont {Lilach}}, \bibinfo {author}
  {\bibfnamefont {O.}~\bibnamefont {Cheshnovsky}},\ and\ \bibinfo {author}
  {\bibfnamefont {M.}~\bibnamefont {Arndt}},\ }\bibfield  {title} {\bibinfo
  {title} {In search of multipath interference using large molecules},\ }\href
  {https://doi.org/10.1126/sciadv.1602478} {\bibfield  {journal} {\bibinfo
  {journal} {Sci. Adv.}\ }\textbf {\bibinfo {volume} {3}},\ \bibinfo {pages}
  {e1602478} (\bibinfo {year} {2017})}\BibitemShut {NoStop}%
\bibitem [{\citenamefont {Johansson}\ \emph {et~al.}(2012)\citenamefont
  {Johansson}, \citenamefont {Nation},\ and\ \citenamefont
  {Nori}}]{NoriCPC2012}%
  \BibitemOpen
  \bibfield  {author} {\bibinfo {author} {\bibfnamefont {J.}~\bibnamefont
  {Johansson}}, \bibinfo {author} {\bibfnamefont {P.}~\bibnamefont {Nation}},\
  and\ \bibinfo {author} {\bibfnamefont {F.}~\bibnamefont {Nori}},\ }\bibfield
  {title} {\bibinfo {title} {{QuTiP}: An open-source python framework for the
  dynamics of open quantum systems},\ }\href
  {https://doi.org/https://doi.org/10.1016/j.cpc.2012.02.021} {\bibfield
  {journal} {\bibinfo  {journal} {Comput. Phys. Commun.}\ }\textbf {\bibinfo
  {volume} {183}},\ \bibinfo {pages} {1760 } (\bibinfo {year}
  {2012})}\BibitemShut {NoStop}%
\bibitem [{\citenamefont {Johansson}\ \emph {et~al.}(2013)\citenamefont
  {Johansson}, \citenamefont {Nation},\ and\ \citenamefont
  {Nori}}]{NoriCPC2013}%
  \BibitemOpen
  \bibfield  {author} {\bibinfo {author} {\bibfnamefont {J.}~\bibnamefont
  {Johansson}}, \bibinfo {author} {\bibfnamefont {P.}~\bibnamefont {Nation}},\
  and\ \bibinfo {author} {\bibfnamefont {F.}~\bibnamefont {Nori}},\ }\bibfield
  {title} {\bibinfo {title} {{QuTiP} 2: A python framework for the dynamics of
  open quantum systems},\ }\href
  {https://doi.org/https://doi.org/10.1016/j.cpc.2012.11.019} {\bibfield
  {journal} {\bibinfo  {journal} {Comput. Phys. Commun.}\ }\textbf {\bibinfo
  {volume} {184}},\ \bibinfo {pages} {1234 } (\bibinfo {year}
  {2013})}\BibitemShut {NoStop}%
\bibitem [{\citenamefont {Brand}\ \emph {et~al.}(2020)\citenamefont {Brand},
  \citenamefont {Simonovi\'c}, \citenamefont {Kia{\l}ka}, \citenamefont
  {Troyer}, \citenamefont {Geyer},\ and\ \citenamefont
  {Arndt}}]{Brand_OptEx2019}%
  \BibitemOpen
  \bibfield  {author} {\bibinfo {author} {\bibfnamefont {C.}~\bibnamefont
  {Brand}}, \bibinfo {author} {\bibfnamefont {K.}~\bibnamefont {Simonovi\'c}},
  \bibinfo {author} {\bibfnamefont {F.}~\bibnamefont {Kia{\l}ka}}, \bibinfo
  {author} {\bibfnamefont {S.}~\bibnamefont {Troyer}}, \bibinfo {author}
  {\bibfnamefont {P.}~\bibnamefont {Geyer}},\ and\ \bibinfo {author}
  {\bibfnamefont {M.}~\bibnamefont {Arndt}},\ }\bibfield  {title} {\bibinfo
  {title} {A fiber-based beam profiler for high-power laser beams in confined
  spaces and ultra-high vacuum},\ }\href {https://doi.org/10.1364/OE.387650}
  {\bibfield  {journal} {\bibinfo  {journal} {Opt. Express}\ }\textbf {\bibinfo
  {volume} {28}},\ \bibinfo {pages} {6164} (\bibinfo {year}
  {2020})}\BibitemShut {NoStop}%
\bibitem [{\citenamefont {Lin}\ and\ \citenamefont {Wu}(2014)}]{LIN20141}%
  \BibitemOpen
  \bibfield  {author} {\bibinfo {author} {\bibfnamefont {C.-C.}\ \bibnamefont
  {Lin}}\ and\ \bibinfo {author} {\bibfnamefont {M.-S.}\ \bibnamefont {Wu}},\
  }\bibfield  {title} {\bibinfo {title} {Degradation of ciprofloxacin by
  {UV/S2O82}-process in a large photoreactor},\ }\href
  {https://doi.org/https://doi.org/10.1016/j.jphotochem.2014.04.002} {\bibfield
   {journal} {\bibinfo  {journal} {J. Photochem. Photobiol. A}\ }\textbf
  {\bibinfo {volume} {285}},\ \bibinfo {pages} {1 } (\bibinfo {year}
  {2014})}\BibitemShut {NoStop}%
\bibitem [{\citenamefont {Ramprasad}\ and\ \citenamefont
  {Shi}(2006)}]{Ramprasad_ApplPhysLett88_222903}%
  \BibitemOpen
  \bibfield  {author} {\bibinfo {author} {\bibfnamefont {R.}~\bibnamefont
  {Ramprasad}}\ and\ \bibinfo {author} {\bibfnamefont {N.}~\bibnamefont
  {Shi}},\ }\bibfield  {title} {\bibinfo {title} {Polarizability of
  phthalocyanine based molecular systems: A first-principles electronic
  structure study},\ }\href@noop {} {\bibfield  {journal} {\bibinfo  {journal}
  {Appl. Phys. Lett.}\ }\textbf {\bibinfo {volume} {88}},\ \bibinfo {pages}
  {222903} (\bibinfo {year} {2006})}\BibitemShut {NoStop}%
\bibitem [{\citenamefont {Du}\ \emph {et~al.}(1998)\citenamefont {Du},
  \citenamefont {Fuh}, \citenamefont {Li}, \citenamefont {Corkan},\ and\
  \citenamefont {Lindsey}}]{Du_PhotochemPhotobiol68_141}%
  \BibitemOpen
  \bibfield  {author} {\bibinfo {author} {\bibfnamefont {H.}~\bibnamefont
  {Du}}, \bibinfo {author} {\bibfnamefont {R.-C.~A.}\ \bibnamefont {Fuh}},
  \bibinfo {author} {\bibfnamefont {J.}~\bibnamefont {Li}}, \bibinfo {author}
  {\bibfnamefont {L.~A.}\ \bibnamefont {Corkan}},\ and\ \bibinfo {author}
  {\bibfnamefont {J.~S.}\ \bibnamefont {Lindsey}},\ }\bibfield  {title}
  {\bibinfo {title} {Photochemcad: A computer-aided design and research tool in
  photochemistry},\ }\href@noop {} {\bibfield  {journal} {\bibinfo  {journal}
  {Photochem. Photobiol.}\ }\textbf {\bibinfo {volume} {68}},\ \bibinfo {pages}
  {141} (\bibinfo {year} {1998})}\BibitemShut {NoStop}%
\bibitem [{\citenamefont {Raman}\ and\ \citenamefont
  {Nagendra~Nath}(1936)}]{NathPIASA1936}%
  \BibitemOpen
  \bibfield  {author} {\bibinfo {author} {\bibfnamefont {C.~V.}\ \bibnamefont
  {Raman}}\ and\ \bibinfo {author} {\bibfnamefont {N.~S.}\ \bibnamefont
  {Nagendra~Nath}},\ }\bibfield  {title} {\bibinfo {title} {The diffraction of
  light by high frequency sound waves: {{Part IV}}},\ }\href
  {https://doi.org/10.1007/BF03046242} {\bibfield  {journal} {\bibinfo
  {journal} {Proc. Indian Acad. Sci.}\ }\textbf {\bibinfo {volume} {3}},\
  \bibinfo {pages} {119} (\bibinfo {year} {1936})}\BibitemShut {NoStop}%
\bibitem [{\citenamefont {Talbot}(1836)}]{Talbot1836}%
  \BibitemOpen
  \bibfield  {author} {\bibinfo {author} {\bibfnamefont {H.~F.}\ \bibnamefont
  {Talbot}},\ }\bibfield  {title} {\bibinfo {title} {{LXXVI}. facts relating to
  optical science. {No. IV}},\ }\href
  {https://doi.org/10.1080/14786443608649032} {\bibfield  {journal} {\bibinfo
  {journal} {Lond. Edinb. Dubl. Phil. Mag.}\ }\textbf {\bibinfo {volume} {9}},\
  \bibinfo {pages} {401} (\bibinfo {year} {1836})}\BibitemShut {NoStop}%
\bibitem [{\citenamefont {M{\"u}ller}\ \emph {et~al.}(2008)\citenamefont
  {M{\"u}ller}, \citenamefont {Chiow},\ and\ \citenamefont {Chu}}]{ChuPRA2008}%
  \BibitemOpen
  \bibfield  {author} {\bibinfo {author} {\bibfnamefont {H.}~\bibnamefont
  {M{\"u}ller}}, \bibinfo {author} {\bibfnamefont {S.-w.}\ \bibnamefont
  {Chiow}},\ and\ \bibinfo {author} {\bibfnamefont {S.}~\bibnamefont {Chu}},\
  }\bibfield  {title} {\bibinfo {title} {Atom-wave diffraction between the
  {{Raman}}-{{Nath}} and the {{Bragg}} regime: {{Effective Rabi}} frequency,
  losses, and phase shifts},\ }\href
  {https://doi.org/10.1103/PhysRevA.77.023609} {\bibfield  {journal} {\bibinfo
  {journal} {Phys. Rev. A}\ }\textbf {\bibinfo {volume} {77}},\ \bibinfo
  {pages} {023609} (\bibinfo {year} {2008})}\BibitemShut {NoStop}%
\bibitem [{\citenamefont {Keller}\ \emph {et~al.}(1999)\citenamefont {Keller},
  \citenamefont {Schmiedmayer}, \citenamefont {Zeilinger}, \citenamefont
  {Nonn}, \citenamefont {D{\"u}rr},\ and\ \citenamefont
  {Rempe}}]{RempeAPB1999}%
  \BibitemOpen
  \bibfield  {author} {\bibinfo {author} {\bibfnamefont {C.}~\bibnamefont
  {Keller}}, \bibinfo {author} {\bibfnamefont {J.}~\bibnamefont
  {Schmiedmayer}}, \bibinfo {author} {\bibfnamefont {A.}~\bibnamefont
  {Zeilinger}}, \bibinfo {author} {\bibfnamefont {T.}~\bibnamefont {Nonn}},
  \bibinfo {author} {\bibfnamefont {S.}~\bibnamefont {D{\"u}rr}},\ and\
  \bibinfo {author} {\bibfnamefont {G.}~\bibnamefont {Rempe}},\ }\bibfield
  {title} {\bibinfo {title} {Adiabatic following in standing-wave diffraction
  of atoms},\ }\href {https://doi.org/10.1007/s003400050810} {\bibfield
  {journal} {\bibinfo  {journal} {Appl. Phys. B}\ }\textbf {\bibinfo {volume}
  {69}},\ \bibinfo {pages} {303} (\bibinfo {year} {1999})}\BibitemShut
  {NoStop}%
\bibitem [{\citenamefont {Ryytty}\ and\ \citenamefont
  {Kaivola}(2000)}]{KaivolaPRL2000}%
  \BibitemOpen
  \bibfield  {author} {\bibinfo {author} {\bibfnamefont {P.}~\bibnamefont
  {Ryytty}}\ and\ \bibinfo {author} {\bibfnamefont {M.}~\bibnamefont
  {Kaivola}},\ }\bibfield  {title} {\bibinfo {title} {Pulsed standing-wave
  mirror for neutral atoms and molecules},\ }\href
  {https://doi.org/10.1103/PhysRevLett.84.5074} {\bibfield  {journal} {\bibinfo
   {journal} {Phys. Rev. Lett.}\ }\textbf {\bibinfo {volume} {84}},\ \bibinfo
  {pages} {5074} (\bibinfo {year} {2000})}\BibitemShut {NoStop}%
\bibitem [{\citenamefont {Savitzky}\ and\ \citenamefont
  {Golay}(1964)}]{Savitzky_AC36_1627}%
  \BibitemOpen
  \bibfield  {author} {\bibinfo {author} {\bibfnamefont {A.}~\bibnamefont
  {Savitzky}}\ and\ \bibinfo {author} {\bibfnamefont {M.~J.~E.}\ \bibnamefont
  {Golay}},\ }\bibfield  {title} {\bibinfo {title} {Smoothing and
  differentiation of data by simplified least squares procedures.},\ }\href
  {https://doi.org/10.1021/ac60214a047} {\bibfield  {journal} {\bibinfo
  {journal} {Anal. Chem.}\ }\textbf {\bibinfo {volume} {36}},\ \bibinfo {pages}
  {1627} (\bibinfo {year} {1964})}\BibitemShut {NoStop}%
\bibitem [{\citenamefont {Shull}(1968)}]{Shull_PhysRevLett21_1585}%
  \BibitemOpen
  \bibfield  {author} {\bibinfo {author} {\bibfnamefont {C.~G.}\ \bibnamefont
  {Shull}},\ }\bibfield  {title} {\bibinfo {title} {Observation of
  pendell\"osung fringe structure in neutron diffraction},\ }\href@noop {}
  {\bibfield  {journal} {\bibinfo  {journal} {Phys. Rev. Lett.}\ }\textbf
  {\bibinfo {volume} {21}},\ \bibinfo {pages} {1585} (\bibinfo {year}
  {1968})}\BibitemShut {NoStop}%
\bibitem [{\citenamefont {Oberthaler}\ \emph {et~al.}(1999)\citenamefont
  {Oberthaler}, \citenamefont {Abfalterer}, \citenamefont {Bernet},
  \citenamefont {Keller}, \citenamefont {Schmiedmayer},\ and\ \citenamefont
  {Zeilinger}}]{Oberthaler_PRA60_456}%
  \BibitemOpen
  \bibfield  {author} {\bibinfo {author} {\bibfnamefont {M.~K.}\ \bibnamefont
  {Oberthaler}}, \bibinfo {author} {\bibfnamefont {R.}~\bibnamefont
  {Abfalterer}}, \bibinfo {author} {\bibfnamefont {S.}~\bibnamefont {Bernet}},
  \bibinfo {author} {\bibfnamefont {C.}~\bibnamefont {Keller}}, \bibinfo
  {author} {\bibfnamefont {J.}~\bibnamefont {Schmiedmayer}},\ and\ \bibinfo
  {author} {\bibfnamefont {A.}~\bibnamefont {Zeilinger}},\ }\bibfield  {title}
  {\bibinfo {title} {Dynamical diffraction of atomic matter waves by crystals
  of light},\ }\href@noop {} {\bibfield  {journal} {\bibinfo  {journal} {Phys.
  Rev. A}\ }\textbf {\bibinfo {volume} {60}} (\bibinfo {year}
  {1999})}\BibitemShut {NoStop}%
\bibitem [{\citenamefont {Piskorski}\ \emph {et~al.}(2014)\citenamefont
  {Piskorski}, \citenamefont {Patterson}, \citenamefont {Eibenberger},\ and\
  \citenamefont {Doyle}}]{DoyleCPC2014}%
  \BibitemOpen
  \bibfield  {author} {\bibinfo {author} {\bibfnamefont {J.}~\bibnamefont
  {Piskorski}}, \bibinfo {author} {\bibfnamefont {D.}~\bibnamefont
  {Patterson}}, \bibinfo {author} {\bibfnamefont {S.}~\bibnamefont
  {Eibenberger}},\ and\ \bibinfo {author} {\bibfnamefont {J.~M.}\ \bibnamefont
  {Doyle}},\ }\bibfield  {title} {\bibinfo {title} {Cooling, spectroscopy and
  non-sticking of trans-stilbene and nile red},\ }\href
  {https://doi.org/10.1002/cphc.201402502} {\bibfield  {journal} {\bibinfo
  {journal} {Chem. Phys. Chem.}\ }\textbf {\bibinfo {volume} {15}},\ \bibinfo
  {pages} {3800} (\bibinfo {year} {2014})}\BibitemShut {NoStop}%
\bibitem [{\citenamefont {Ben~Dahan}\ \emph {et~al.}(1996)\citenamefont
  {Ben~Dahan}, \citenamefont {Peik}, \citenamefont {Reichel}, \citenamefont
  {Castin},\ and\ \citenamefont {Salomon}}]{SalomonPRL1996}%
  \BibitemOpen
  \bibfield  {author} {\bibinfo {author} {\bibfnamefont {M.}~\bibnamefont
  {Ben~Dahan}}, \bibinfo {author} {\bibfnamefont {E.}~\bibnamefont {Peik}},
  \bibinfo {author} {\bibfnamefont {J.}~\bibnamefont {Reichel}}, \bibinfo
  {author} {\bibfnamefont {Y.}~\bibnamefont {Castin}},\ and\ \bibinfo {author}
  {\bibfnamefont {C.}~\bibnamefont {Salomon}},\ }\bibfield  {title} {\bibinfo
  {title} {Bloch oscillations of atoms in an optical potential},\ }\href
  {https://doi.org/10.1103/PhysRevLett.76.4508} {\bibfield  {journal} {\bibinfo
   {journal} {Phys. Rev. Lett.}\ }\textbf {\bibinfo {volume} {76}},\ \bibinfo
  {pages} {4508} (\bibinfo {year} {1996})}\BibitemShut {NoStop}%
\end{thebibliography}
%

\end{document}


\title{Supplemental Material for \\ Bragg diffraction of large organic molecules}

\author{Christian Brand}
\affiliation{University of Vienna, Faculty of Physics, Boltzmanngasse 5, A-1090 Vienna, Austria}
\affiliation{German Aerospace Center (DLR), Institute of Quantum Technologies, Söflinger Stra\ss e 100, 89077 Ulm, Germany}

\author{Filip Kia{\l}ka}
\affiliation{University of Vienna, Faculty of Physics, Boltzmanngasse 5, A-1090 Vienna, Austria}
\affiliation{Faculty of Physics, University of Duisburg-Essen, Lotharstra\ss e 1, 47048 Duisburg, Germany}

\author{Stephan Troyer}
\affiliation{University of Vienna, Faculty of Physics, Boltzmanngasse 5, A-1090 Vienna, Austria}

\author{Christian Knobloch}
\affiliation{University of Vienna, Faculty of Physics, Boltzmanngasse 5, A-1090 Vienna, Austria}

\author{Ksenija Simonovi\'c}
\affiliation{University of Vienna, Faculty of Physics, Boltzmanngasse 5, A-1090 Vienna, Austria}

\author{Benjamin A. Stickler}
\affiliation{Faculty of Physics, University of Duisburg-Essen, Lotharstra\ss e 1, 47048 Duisburg, Germany}
\affiliation{QOLS, Blackett Laboratory, Imperial College London, SW7 2AZ London, United Kingdom}

\author{Klaus Hornberger}
\affiliation{Faculty of Physics, University of Duisburg-Essen, Lotharstra\ss e 1, 47048 Duisburg, Germany}

\author{Markus Arndt}
\email{markus.arndt@univie.ac.at}
\affiliation{University of Vienna, Faculty of Physics, Boltzmanngasse 5, A-1090 Vienna, Austria}

\date{\today}

\maketitle

\section{Laser Desorption}

	We employ a tightly focused laser beam to thermally evaporate the molecules.
	This results in a high thermal load which may lead to thermal decomposition of ciprofloxacin, especially the detachment of the carboxyl group (-COOH).
	As the experimental setup offers no mass resolution, this might deteriorate the contrast of the observed pattern.
	To test whether fragmentation occurs in our source, we desorbed ciprofloxacin in high vacuum and collected the material 8~mm behind the source.
	This sample was analyzed using matrix-assisted laser desorption/ionization mass spectrometry and compared to the pristine sample from the supplier.
	The mass spectra show that the laser evaporation leaves more than 99\% of ciprofloxacin intact.
	For phthalocyanine this has been tested with the same result~\cite{PhD_ChristianKnobloch}.

	\begin{figure}
		\includegraphics[width=.9\columnwidth]{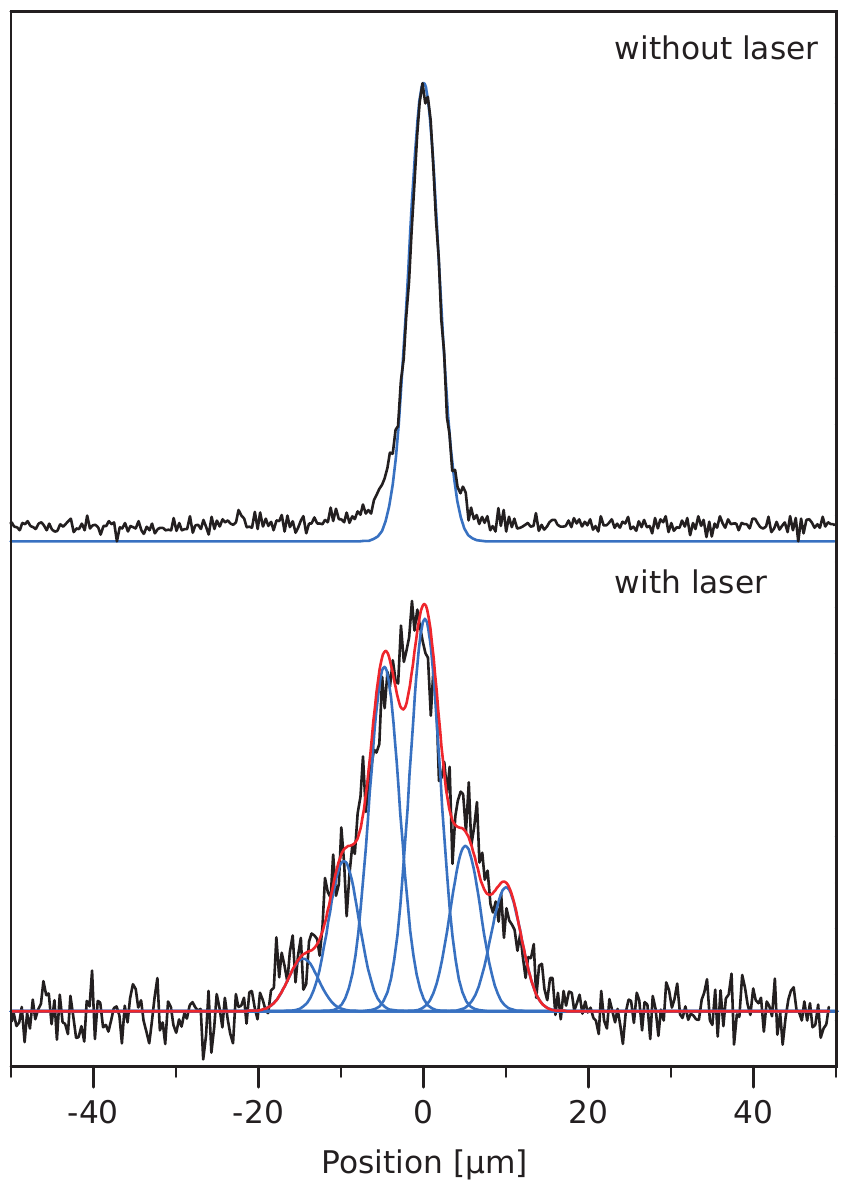}
		\caption{
			a) Collimating a beam of phthalocyanine with the S\textsubscript{x} delimiter set to $4 \: \text{µm}$ leads to a Gaussian signal with a $1/e^2$ radius of $w_x=4.4(1)~\text{µm}$ at the detector.
			b) Inside a $30 \: \text{W}$ laser beam the molecules absorb a mean number of 2 to 3 photons for $v=140$~m/s resulting in a broadened pattern.
			The spacing of the peaks ($4.9~\text{µm}$) matches the recoil of a single 532~nm photon and the resolved substructure suggests that re-emission after absorption is not the dominant deexcitation mechanism.
		}
		\label{fig:Absorption}
	\end{figure}

\section{Fluorescence imaging} 
\label{sec:fluorescence_imaging}

	To visualize the diffraction images of ciprofloxacin, we illuminate the pattern with about 100~mW of 266~nm light generated by a \textsc{Sirah Wavetrain 2} pumped by a \textsc{Coherent Verdi V10}.
	We use a rotating diffuser to achieve uniform illumination at a grazing angle of incidence.
	The fluorescence photons are collected via a 20-fold microscope objective (\textsc{Zeiss PlanNeo Fluar}, $\text{NA}=0.5$) and separated from the background via a bandpass filter transmitting light in the range between 505 and 595 nm.
	The images are recorded with a UV enhanced EMCCD camera (\textsc{Andor iXon DV885 - K(S-VP)}), using a multiplication factor of 1 and an integration time of 20~s.
	Background-correction was achieved by subtracting images under identical illumination with and without molecules.

	The patterns of phthalocyanine are recorded by illuminating the pattern with 661~nm light and recording the fluorescence in the range between 700 and 725 nm.
	For more details see Refs.~\cite{Juffmann2012, ArndtSA2017}.


\section{Data processing --- Ciprofloxacin} 
\label{sec:data_processing_ciprofloxacin}

	We perform data processing of all diffraction images using the \textsc{SciPy} stack.
	For Fig.~\ref{MT-fig:Bragg_cipro}a) we averaged 6 individual images of the deposited pattern and denoised the result with a Gaussian filter with a radius of 1 camera pixel.
	We then perform background correction (in addition to the background subtraction done after image acquisition) by masking the diffraction pattern, averaging the image along the $y$-direction, fitting a smoothing spline, and subtracting the noise floor.
	The same process is repeated along the $x$-axis.
	To find the horizontal center of the diffraction pattern we fit a Gaussian to a $y$-averaged, $16 \: \text{µm}$ wide horizontal stripe at the top of the diffraction pattern.


\section{Data processing --- Phthalocyanine} 
\label{sec:data_processsing_phthalocyanine}

	We determine the forward velocities in the images by comparison with a phthalocyanine diffraction pattern obtained with a material grating in Ref.~\cite{ArndtSA2017}.
	The patterns are aligned by maximizing the overlap of their intensity distributions, which are obtained by integrating the images horizontally.
	For a material grating the diffraction orders are clearly separated, and the position- as well as momentum-space separation between them is known, which allows us to calculate the forward velocities.

	To align the profiles in Fig.~\ref{MT-fig:RockingCurve}d) we fit them with a sum of three Gaussians, two narrow ones for the peaks and a broad one to account for the losses.
	The profiles and images are then horizontally aligned with respect to the rightmost Gaussian for negative incidence angles and the leftmost Gaussian for positive incidence angles.
	By taking into account the molecules' forward velocity, we convert the horizontal axis from pixel to $\hbar k$.


\section{Numerical simulation --- Ciprofloxacin} 
\label{sec:numerical_simulation}

	The diffraction image shown in Fig.~2b) is simulated line-by-line (horizontally) by solving the Raman-Nath equations~\eqref{MT-eq:Raman-Nath} using \textsc{QuTiP}~\cite{NoriCPC2012,NoriCPC2013}.
	We truncate the infinite set of equations to those with $\abs{j} < 2^{13}$ and choose $n = 700$.
	The initial state is Gaussian in position space with a parabolic phase and a standard deviation of the probability amplitude equal to $4.6 \: \text{µm}$.
	The latter is chosen so that the width of the undiffracted beam at the top of the experimental image matches that in the simulation with the laser turned off.
	The parabolic phase, in turn, is that of a paraxially-approximated spherical wave with the source located $1505 \: \mathrm{mm}$ (the distance between the source and the S\textsubscript{x} delimiter) away.
	We start by transforming the initial state to momentum space via FFT and evolving it using \textsc{QuTiP}'s \verb!sesolve! with a time-dependent, band-diagonal Hamiltonian.
	We integrate the Schrödinger equation over a time interval of $6 \sigma \omega_r^{-1}$, after which free propagation in momentum space (by multiplication with the transfer function in Fresnel approximation) is performed.
	The result is then transformed back to position space.
	The resulting lines of the simulation are stacked vertically and multiplied by the intensity of the corresponding data line.
	Then, a Gaussian filter is applied in the vertical direction to the obtained image to account for the finite height of S\textsubscript{y}.
	To account for the horizontal extent of the source (and thus finite transverse coherence), we calculate 50 diffraction patterns for point sources with different $x$ positions and average the images by intensity with Gaussian weights corresponding to an estimated source radius of $12 \: \text{µm}$ standard deviation.


\section{Absorption inside the grating} 
\label{sec:absorption}

	To estimate the number of photons phthalocyanine absorbs inside the laser grating, we limit the transverse velocity spread in the molecular beam to about the recoil velocity by closing the S\textsubscript{x} delimiter to $4 \: \text{µm}$.
	For molecules traveling at 140~m/s, this leads to a most probable transverse velocity of 0.4~mm/s, which corresponds to a kinetic energy of about 5~nK in this degree of freedom.
	With the laser grating turned off, the signal at the detector has a 1/e$^2$ radius of $4.4(1)~\text{µm}$ as shown in Suppl. Fig.~\ref{fig:Absorption}a).
	Turning on the grating with a power of $30 \: \text{W}$, vertical radius $w_y=44(1) \: \text{µm}$, and incidence angle $\theta_{\rm grat}=1.25 \: \text{mrad}$, for which we expect no diffraction, results in a broadening of the beam as shown in Suppl. Fig.~\ref{fig:Absorption}b).
	The lineshape exhibits a substructure whose spacing  matches the recoil of a 532~nm photon for molecules travelling at 140~m/s, assuming that the width of the individual peaks remains constant.
	From the shape we infer that the mean number of absorbed photons is about $2.5$ at this laser intensity, and thus in the range $0.8$--$1.0$ at the intensities used in the diffraction experiments.
	

	\begin{figure}
		\includegraphics[width=.6\columnwidth]{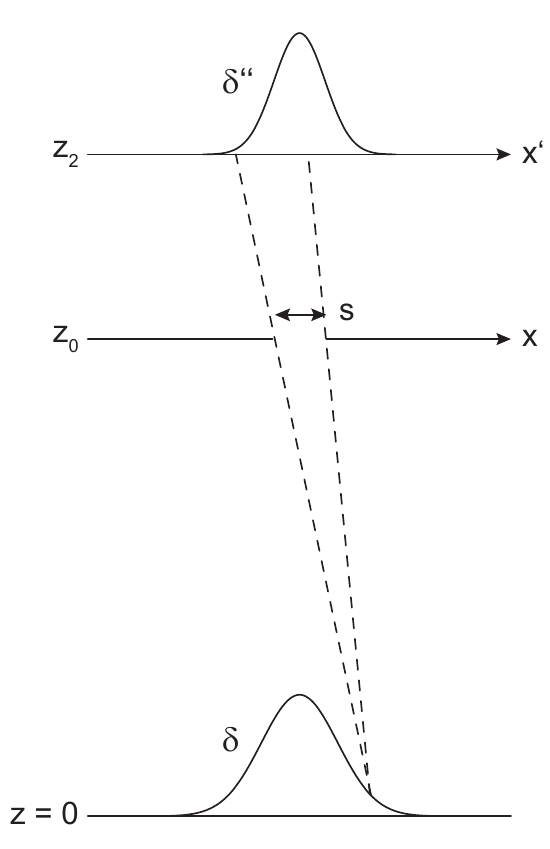}
		\caption{
			Estimating the effective source size $\delta$ and the collimation radius of the molecular beam from the known on-screen stripe radius $\delta^{\prime \prime}$ and collimation slit width $s$.
			The Gaussian peaks represent the (approximately Gaussian) molecular densities in the source and detector planes.
		}
		\label{fig:source_size}
	\end{figure}

	\begin{figure}
		\includegraphics[width=.9\columnwidth]{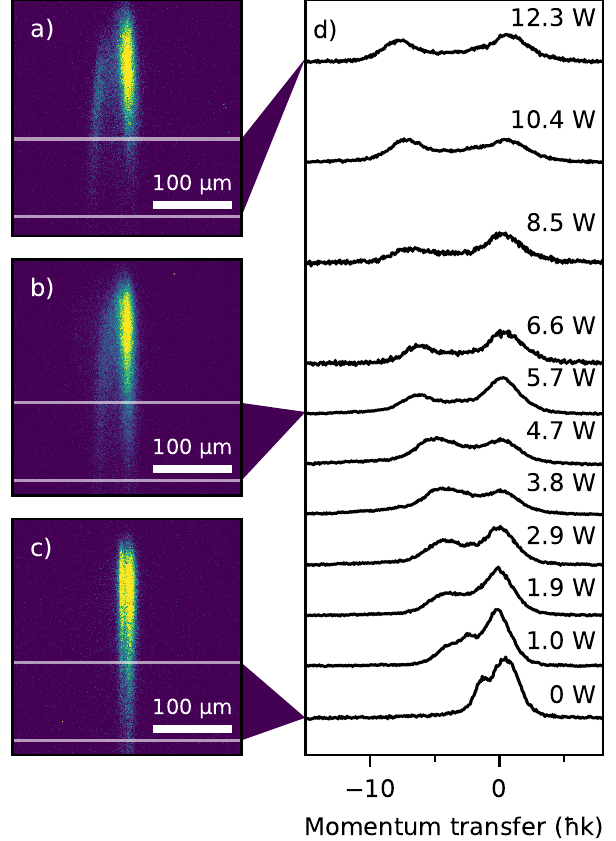}
		\caption{
			Power dependence of Bragg diffraction of phthalocyanine.
			Panels (a) and (b) show the diffraction patterns for laser grating powers of 12.3 (a) and 5.7~W (b). 
			The double peak visible at 0~W (c) is an artifact caused by the collimation slit.
			Panel (d) shows intensity profiles integrated over a region corresponding to a velocity range of 143--175~m/s.
		}
		\label{fig:powerscan}
	\end{figure}

\section{Molecular beam collimation} 
\label{sec:collimation}

	To estimate the collimation radius of the molecular beam, we first estimate the source size using ray optics, as illustrated in Suppl. Fig.~\ref{fig:source_size}.
	For an infinitely narrow collimation slit at $z = z_0$ and Gaussian source (at $z = 0$) with standard deviation $\delta$, we would expect a Gaussian stripe on screen with width
	\begin{equation}
		\delta^{\prime} = \frac{z_2 - z_0}{z_0} \delta.
	\end{equation}
	If the slit has finite width described by a transmission function $t(x)$, the stripe on screen will be a convolution of the $\delta^{\prime}$-wide Gaussian with a projection of the slit, $t\left((z_0/z_2) x'\right)$.
	To obtain a simple analytical estimate of the stripe width, we approximate a boxcar-shaped $t(x)$ with a Gaussian with a standard deviation of $s/4$.
	The stripe is then also Gaussian with a standard deviation
	\begin{equation} \label{eq:stripe_width}
		\delta^{\prime \prime} = \sqrt{\left(\frac{s z_2}{4 z_0}\right)^2 + \left(\frac{z_2 - z_0}{z_0} \delta\right)^2}.
	\end{equation}
	Eq.~\eqref{eq:stripe_width} is easily inverted, allowing us to estimate $\delta$ knowing $s$ and $\delta^{\prime \prime}$.
	With a known source size $\delta$ and slit size $s$ we can estimate the one-sigma collimation radius to be
	\begin{equation}
		\frac1{z_0} \left(\delta + \frac{s}{2}\right).
	\end{equation}

	Using Eq.~\eqref{eq:stripe_width} we estimate the source sizes to be $12$ and $10 \: \text{µm}$ for the data in Figs.~\ref{MT-fig:pch2_pendelloesung} and~\ref{MT-fig:RockingCurve}, respectively.
	This gives one-sigma collimation radii of $12 \: \text{µrad}$ in both cases (the two-sigma collimation radii are $20$ and $19 \: \text{µrad}$, respectively).


\section{Effect of grating power} 
\label{sec:power_scan}

	To study the influence of the potential depth on the diffraction efficiency, we record diffraction patterns at grating powers ranging from 0 up to 12.3~W, as shown in Suppl. Fig.~\ref{fig:powerscan}.
	The profiles in panel (d) show the intensity oscillating between the diffracted and the undiffracted beams.
	Additionally, the distance between the peaks increases as a function of power, corresponding to a change in $\theta_{\rm{grat}}$ of about 20~µrad. 
	We attribute this to residual thermal drift in the experimental setup.


%